\documentclass[%
 reprint,
superscriptaddress,
 amsmath,amssymb,
 aps,
floatfix,
]{revtex4-2}
\usepackage{comment}
\usepackage{graphicx}% Include figure files
\usepackage{array}[=2016-10-06]
\usepackage{dcolumn}% Align table columns on decimal point
\usepackage{bm}% bold math
\usepackage{todonotes}

\usepackage{hyperref}% add hypertext capabilities
\usepackage{booktabs}

\usepackage{enumitem}
\usepackage{titlesec}
\usepackage{diagbox}

\titleformat{\section}
  {\normalfont\bfseries\scshape\centering}
  {\thesection}{1em}{}

\titlespacing*{\subsection}
{0pt}{1.5ex plus 1ex minus .2ex}{1ex plus .2ex}
\titlespacing*{\subsubsection}
{0pt}{1.5ex plus 1ex minus .2ex}{1ex plus .2ex}

\usepackage{array, multirow}
\usepackage{physics}
\usepackage{ragged2e}

\usepackage{caption}
\usepackage{subcaption}
\DeclareCaptionFormat{justifcaption}{%
\begin{minipage}{\linewidth}\noindent\justifying\small #1#2#3\end{minipage}}
\DeclareCaptionFormat{centercaption}{%
\begin{minipage}{\linewidth}\noindent\Centering\small #1#2#3\end{minipage}}
\usepackage{tikz}
\usetikzlibrary{positioning,arrows.meta,calc}
\usepackage{standalone}

\usepackage{accents}
\newcommand{\thickbar}[1]{\accentset{\rule{.6em}{0.8pt}}{#1}}

\makeatletter
\renewcommand{\fnum@figure}{\textbf{Fig.~\thefigure}}  % change label
\makeatother

\newcommand{\demxyz}{D_{XYZ}}
\newcommand{\demx}{D_{X}}
\newcommand{\demz}{D_{Z}}
\newcommand{\demxyzbar}{\thickbar{D}_{XYZ}}

\usepackage{xspace}
\newcommand{\garipp}{GARI\texttt{++}\xspace}

\begin{document}

\title{\garipp: A General-Purpose Decoder for Quantum Error Correction}

\author{Arshpreet Singh Maan}
    \email{arshpreet.maan@aalto.fi}
\affiliation{Aalto University, School of Science, Department of Computer Science, Konemiehentie~2, 02150 Espoo, Finland}

\author{Francisco Garcia-Herrero}
    % \email{francg18@ucm.es}
\affiliation{Universidad Complutense de Madrid, Madrid, Spain}

\author{Daniel Báscones}
\affiliation{Universidad Complutense de Madrid, Madrid, Spain}

\author{Valentin Savin}
% \email{valentin.savin@quobly.io}
%\affiliation{Universit\'e Grenoble Alpes, CEA-L\'eti, Grenoble, France}
\affiliation{Quobly, Grenoble, France}

\author{Alexandru Paler}
 % \email{alexandru.paler@aalto.fi}
\affiliation{Aalto University, School of Science, Department of Computer Science, Konemiehentie~2, 02150 Espoo, Finland}

% \date{}

\begin{abstract}
We present \garipp, a two-stage decoding pipeline achieving accuracy competitive with the state of the art across surface, bivariate bicycle and color codes without relying on ensembles. The first stage uses GARI, a normalized min-sum decoder with random serial scheduling on the GARI transformed matrix. The second stage decoder is used only for the cases that the first stage does not converge, and the decoder is configurable based on the choice of the QEC code. The first stage pipelines towards the second stage both the residual syndrome and the soft information. \garipp pipelining is predictive, meaning that \garipp chooses the first stage iteration outputs which are expected to deliver the best accuracy in the second stage. \garipp motivates a hybrid FPGA-CPU architecture for real-time decoding. For example, under uniform circuit-level depolarising noise with a physical error rate of 0.001, the FPGA implementation of GARI achieves average latencies per round ranging from $201 ns$ to $11.8\mu s$ for surface code distances between 7 and 25. At the same time, due to the very low weight residuals output by the first stage, the average residual decoding time per round remains near constant and below one microsecond on a single Apple M2 core for surface code distances even up to 51 when using Union-Find and accuracy on par with correlated MWPM.

\end{abstract}

\maketitle

\begin{figure*}[!t]
\centering
\begin{minipage}[t]{0.49\textwidth}
\raggedright\textbf{(a)}\par
\centering
\includegraphics[width=\linewidth]{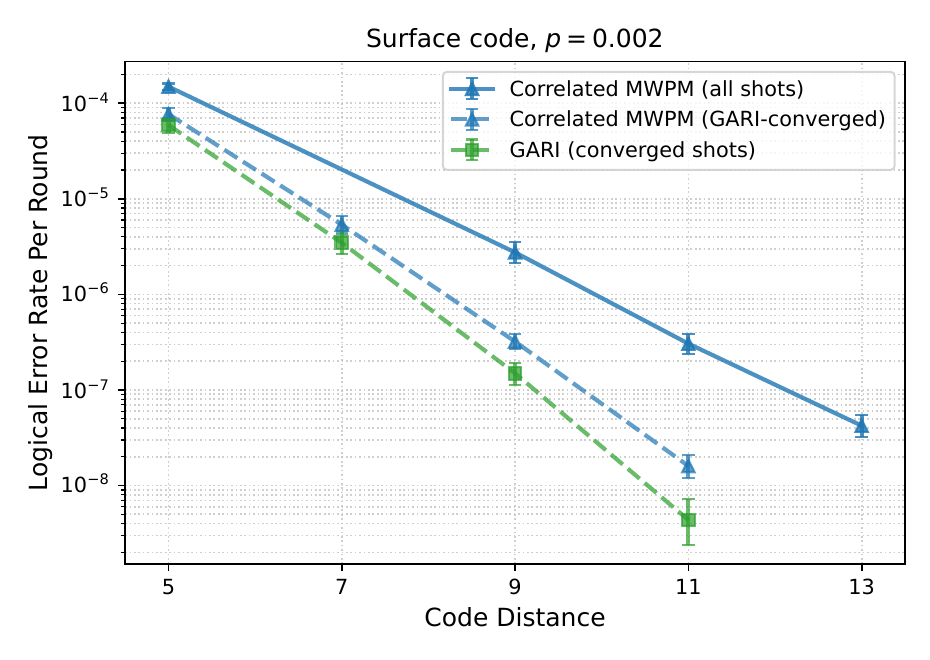}
\end{minipage}\hfill
\begin{minipage}[t]{0.49\textwidth}
\raggedright\textbf{(b)}\par
\centering
\includegraphics[width=\linewidth]{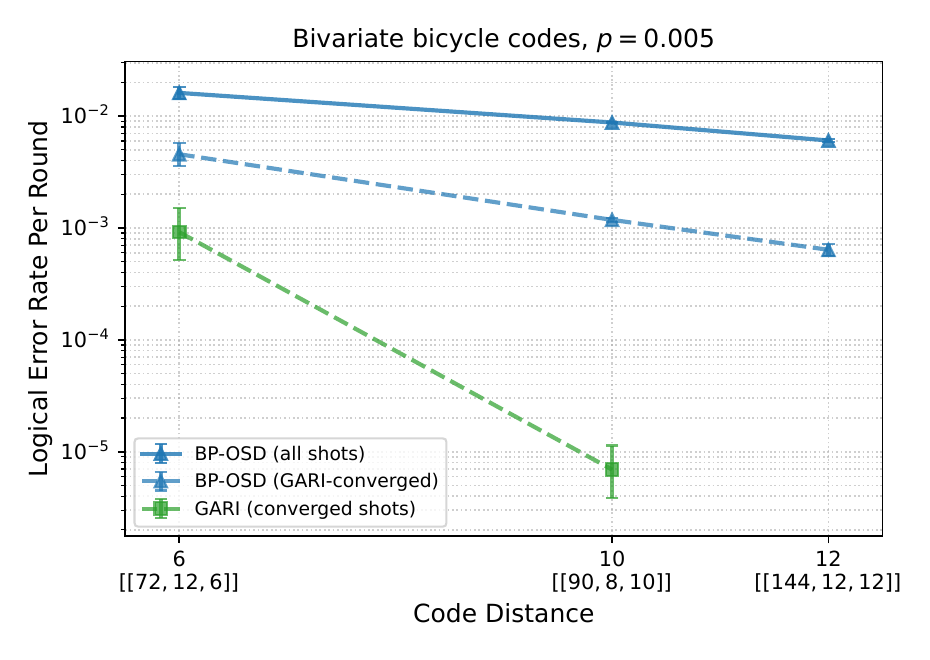}
\end{minipage}
\par\vspace{0.5em}
\begin{minipage}[t]{0.49\textwidth}
\raggedright\textbf{(c)}\par
\centering
\includegraphics[width=\linewidth]{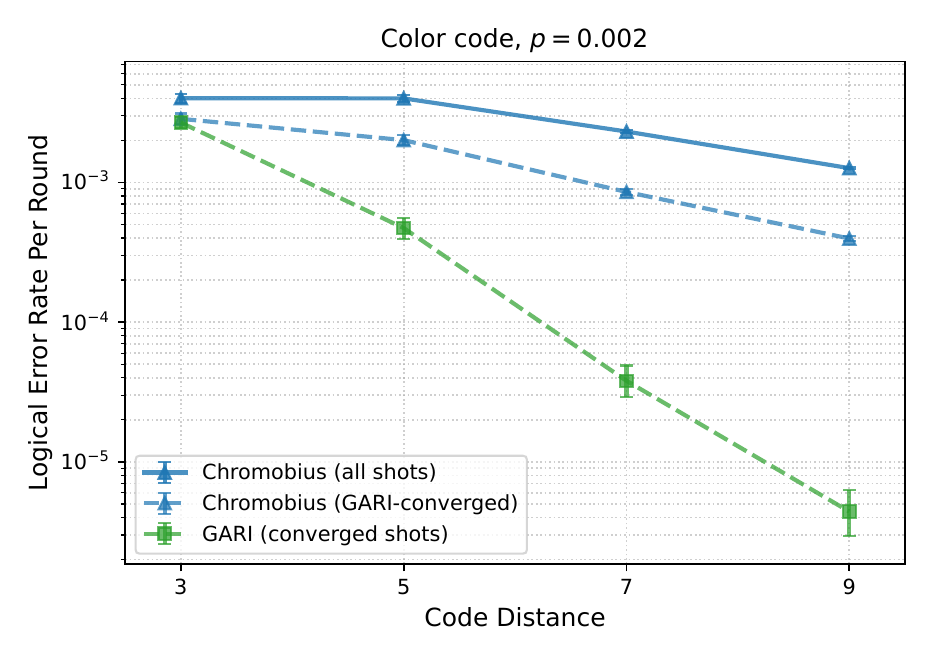}
\end{minipage}
\caption{
Logical error rate per round versus code distance for (a) Surface code at (p=0.002), (b) Bivariate bicycle code at (p=0.005), and (c) Superdense color code at (p=0.002). Blue triangles show the reference decoder correlated MWPM, BP-OSD, and Chromobius, respectively, evaluated over all shots (solid lines) and restricted to GARI-converged shots (dashed lines). Green squares show logical error rates of GARI on the cases it converges.  On those exact shots, GARI exhibits lower logical error rates than the corresponding reference decoder. For each decoder, logical error rate (then converted to per round) for the converged cases is calculated as its number of logical failures among GARI-converged shots divided by the total number of GARI-converged shots.
}
\label{fig:bp_converged_cases}

\end{figure*}

\begin{figure*}[!t]
\centering
\includegraphics[width=\textwidth]{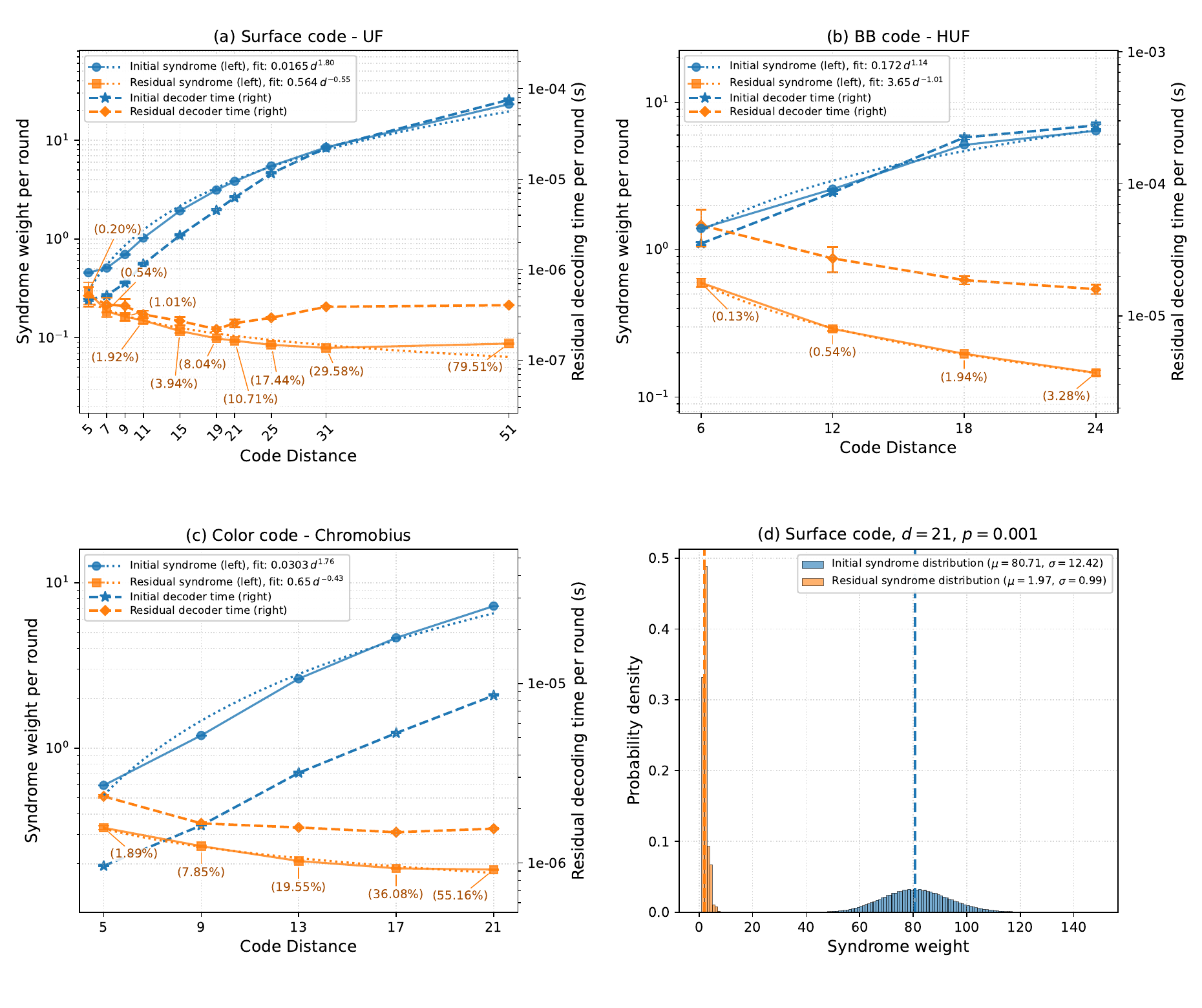}
\caption{Syndrome suppression and decoding time per round under uniform circuit-level depolarizing noise at $p=10^{-3}$ for (a) surface codes with UF, (b) BB codes with HUF, and (c) color codes with Chromobius. Solid blue circles and orange squares show initial and residual $D_Z$ syndrome weights (left axes). Dashed blue stars and orange diamonds show the corresponding initial and residual decoding times (right axes), using original priors and weights computed by GARI, respectively. Parenthesized percentages give the GARI non-convergence fraction over all sampled shots. In (a)--(c), each mean is conditioned on its respective nonzero input syndrome and divided by $d$ rounds; residual samples are restricted to non-converged shots. Dotted curves show fitted lines, with formulas beside the syndrome labels. GARI uses at most seven iterations, with normalization factors $0.9063$, $0.725$, and $0.825$ in (a), (b), and (c). CPU timings use a single Apple M2 core and exclude GARI and decoder/input preparation. (d) Distribution of $Z$-detector syndrome weights for a distance-$21$ surface code at $p=10^{-3}$, before GARI (blue) and after GARI for non-converged shots only (orange). Dashed lines mark means $80.71$ and $1.97$. The residual distribution is concentrated at a few non-zero detectors.}

\label{fig:syndrome_timing}
\label{fig:syndrome_supp_per_round}
\label{fig:syndrome_distribution}
\label{fig:timing_surface_bb}
\label{fig:timing_color}
\label{fig:syndrome_color}
\end{figure*}

\begin{figure*}[!t]
\centering
\includegraphics[width=0.85\textwidth]{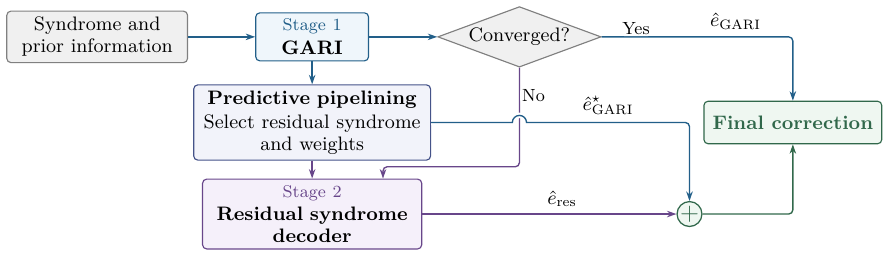}
\caption{\garipp decoding pipeline. Stage~1 applies GARI to the input syndrome and prior information. If GARI converges, its error estimate $\hat e_{\mathrm{GARI}}$ is accepted as the final correction. Predictive pipelining retains a residual syndrome, LLR-derived weights, and the associated partial error estimate $\hat e_{\mathrm{GARI}}^\star=\hat e_{\mathrm{GARI}}^{(t_*)}$ from a selected iteration $t_*$. If convergence is not reached within the iteration budget, Stage~2 decodes the selected residual syndrome on $D_X$ or $D_Z$ using these weights and returns $\hat e_{\mathrm{res}}$. The final correction is $\hat e_{\mathrm{GARI}}^\star\oplus\hat e_{\mathrm{res}}$; the $+$ node denotes XOR. See sec.~\ref{sec:two_stage_decoding} and~\ref{sec:iteration_selection} for details.}
\label{fig:decoding_pipeline}
\end{figure*}

\section{Introduction}
\label{sec:introduction}

Quantum error correction (QEC) is essential for reliable and scalable fault-tolerant quantum computation (FTQC).
Recent advances in quantum hardware ~\cite{google2024quantum,bluvstein2024logical,lacroix2024scaling} have sparked interest in the realisation of experimental QEC protocols.
Moreover, advances in quantum low-density parity-check (QLDPC) codes promise to significantly reduce the qubit overheads for error-corrected computation~\cite{panteleev2022asymptotically,leverrier2022quantum,bravyi2024high,zhao2026ultrahighrate,yang2026gala,bhardwaj2026highrate,higgott2026modular}.
However, the practicality of a QEC code also depends on the decoder used to infer corrections from the measured syndrome. High logical accuracy must be achieved within the timing budget imposed by repeated syndrome extraction and the underlying hardware ~\cite{ryan2021realization,caune2024demonstrating}.

Implementing fast and accurate decoders is a central challenge for realizing fault-tolerant quantum computing (FTQC). For codes with matchable decoding graphs, such as surface codes, minimum weight perfect matching (MWPM) and union-find (UF) provide accurate decoding, at a computational cost that increases with the syndrome weight ~\cite{higgott2025sparse,Delfosse2021almostlineartime}. 

For more general QLDPC codes, a single fault can trigger more than two detection events, resulting in hyperedges in the decoding graph. This requires decoders capable of handling the hypergraph structures~\cite{wu2025minimum,delfosse2022toward}, with computational efficiency and scalability remaining key challenges as the code distance and syndrome weight increase.

Message-passing (MP), also known as belief-pro\-pa\-ga\-tion (BP),  decoders offer a complementary approach. Their local updates and linear complexity per iteration make them attractive for parallel hardware. However, short cycles and quantum degeneracy can prevent such decoders from converging to a useful estimate~\cite{panteleev2021degenerate,raveendran2021trapping,demarti2024decoding,du2025blindness}. Several strategies have been developed to improve their accuracy. One approach is to use BP-derived reliability information to guide a post-processing algorithm, as in BP with ordered-statistics decoding (BP-OSD) and BP with localized statistics decoding (BP-LSD)~\cite{panteleev2021degenerate,roffe2020decoding,hillmann2024localized}. Another is to use the updated BP beliefs to improve the weights of a subsequent graph-based decoder, as in belief-matching and belief-find~\cite{higgott2023improved}. In these approaches, BP guides the subsequent decoding procedure but does not contribute a partial correction that is retained. Alternatively, BP can be used as a partial decoder: even when it does not fully resolve the syndrome, its partial correction is retained and the residual syndrome ~\cite{smith2023localpredecoder} is passed to a conventional decoder. This strategy was proposed by Caune \emph{et al.}, who demonstrated its benefits in combination with MWPM for surface codes~\cite{caune2023belief}. Another strategy is ensemble decoding, which improves decoding performance by combining multiple decoding attempts. This can be achieved, for instance, through automorphism ensembles~\cite{koutsioumpas2025automorphismensembledecodingquantum}, or through information relaying between decoding instances, as in Relay-BP~\cite{muller2025improvedbeliefpropagationsufficient}. While ensembles improve decoding accuracy, they increase computational resources or decoding latency.

A further challenge is to exploit correlations between Pauli error components. Even for CSS codes, decoding the $X$- and $Z$-syndromes independently fails to exploit the information about $Y$-errors, and this leads to suboptimal performance. Correlated (i.e. joint) decoding can account for these correlations, but the resulting correlated detector model exhibits clusters of 4-cycles, which are detrimental to MP decoders. The performance of the MP decoder therefore depends on how these correlations are encoded in the underlying bipartite graph.

Graph Augmentation and Rewiring for Inference (GARI) improves MP decoding by modifying the underlying bipartite graph on which inference is performed~\cite{maan2026decoding}. The resulting GARI graph retains the original $X$- and $Z$-error subgraphs, which are coupled through auxiliary error and check nodes. This construction preserves the joint decoding problem while eliminating clusters of 4-cycles arising from biclique structures associated with $Y$-type errors. High decoding accuracy is achieved using normalized min-sum (NMS) decoding with a hybrid serial-layered schedule: randomized horizontal serial scheduling is applied to the original $X$- and $Z$-error subgraphs, while horizontal layered scheduling is used for the auxiliary part of the graph. Hereafter, we refer to this as the GARI decoder.  

\subsection{Contributions}

We propose \garipp, a two-stage decoding pipeline that combines a single GARI decoder with a second-stage decoder for the residual syndrome. The approach is motivated by two observations about GARI. 

\textbf{Observation 1: } Whenever it converges, GARI achieves a lower logical error rate than conventional decoders. This first observation concerns the quality of the solutions found by GARI alone. The solutions accepted directly from GARI already provide high logical accuracy, without a second decoding stage. Figure~\ref{fig:bp_converged_cases} compares GARI with correlated MWPM on exactly the same GARI-converged surface code shots as correlated MWPM is the standard for benchmarking surface codes. On this subset, GARI achieves a lower logical error rate at all tested distances as shown in Figure~\ref{fig:bp_converged_cases}. 

\textbf{Observation 2: } When GARI does not converge, it leaves only a small residual syndrome while the initial syndrome grows with distance. We finalize the decoding of the residual syndrome using UF for surface codes, hypergraph union-find (HUF) for BB codes, and Chromobius ~\cite{gidney2023newcircuits} for color codes with weights obtained from the same GARI decoder. Figure~\ref{fig:syndrome_timing} shows strong syndrome suppression for surface, bivariate bicycle (BB), and color codes.

While correlated MWPM serves as the accuracy reference for surface codes, the choice of UF for second stage comes from UF being a local algorithm with available hardware implementations ~\cite{liyanage2024fpga, ziad2025local, barber2025realtime}. Moreover, typically UF is less accurate than the MWPM ~\cite{Delfosse2021almostlineartime,higgott2023improved}, here we show that even with using UF as second stage, provided that it receives weights from GARI as first stage, the pipeline achieves the same accuracy as correlated MWPM.

\garipp thus exploits both possible outcomes of the first stage: corrections are accepted directly when GARI converges, while non-converged outputs are used to define a smaller residual decoding problem. Rather than restarting the decoding process from the original syndrome, the second stage builds on the progress made by GARI, using both the residual syndrome and the updated error weights.

Compared to the method of Caune \emph{et al.}~\cite{caune2023belief}, \garipp is introducing three methodological novelties. First, in contrast to~\cite{caune2023belief}, \garipp uses in the first stage the GARI decoder, thereby enabling accurate joint decoding of correlated errors. Second, \garipp uses for the second stage both the residual syndrome and the error weights derived from the first stage, while~\cite{caune2023belief} are discarding the first stage soft information entirely while keeping the residual syndromes. Third, \emph{\garipp uses predictive pipelining}, meaning that \garipp is analysing online the soft information stream generated at each GARI iteration, and decides just in time which first stage iteration would deliver the most promising soft information with respect to decoding accuracy in the second stage. Therefore, while standard two-stage decoders are sending the soft information (i.e. updated error weights) of the last iteration to the second stage, \garipp sends in general the values computed at earlier iterations.

The three methodological novelties (1. GARI for correlated computation of residual syndromes, 2. transmission of soft information and residual syndromes towards the second stage, 3. predictive pipelining of soft information for improved accuracy) enable fast and accurate decoding, and make \garipp a general-purpose decoder. We demonstrate the performance of \garipp across surface, BB, and color codes.

The two-stage decoding architecture is attractive for hardware implementations: local GARI updates can run on an FPGA, while a CPU handles the remaining residual problem. Previous GARI FPGA architectures provide a hardware basis for the first stage~\cite{maan2026decoding,bascones2026scalable}.

The paper is organized as follows. Section~\ref{sec:background} reviews detector error models, the GARI construction, and the decoder used in this work. Section~\ref{sec:methods} introduces \garipp, including the selection of an intermediate GARI state and the residual syndrome and soft information passed to the second stage. Section~\ref{sec:results} presents the complete-decoder accuracy for surface, BB and color codes, examines the residual decoding times and the FPGA implementation of GARI. 

\section{Background}
\label{sec:background}

A detector error model (DEM) represents circuit faults by binary error variables $e$ and a matrix $D$ mapping them to detection events~\cite{Gidney_2021,derks2024designingfaulttolerantcircuitsusing}. A syndrome consistent estimate satisfies $D\hat e=s$, with operations over $\mathbb F_2$, although consistency alone does not guaranty logical success. 

Belief propagation (BP) approximates the marginal probabilities of error variables conditioned on the syndrome by iteratively exchanging messages between variable nodes and parity check nodes~\cite{kschischang2001factor,poulin2008iterative}. Messages and decoder outputs can be represented by log-likelihood ratios (LLRs), defined as,
\begin{equation}
\label{eq:llr_def}
\text{LLR}_i = \log \frac{P(e_i = 0)}{P(e_i = 1)}.
\end{equation}
In the log likelihood ratio (LLR) representation, a variable-to-check message is the prior LLR plus the messages from the other neighboring checks. A check-to-variable message combines the other incident messages subject to the measured parity. The prior LLR and all incoming check messages determine the output LLR for each variable. 

The min-sum (MS) algorithm approximates the nonlinear BP ~\cite{chen2005reduced} check update by a signed minimum: the magnitude is the smallest absolute value among the incoming messages, excluding that from the recipient, and the sign is their sign product multiplied by $(-1)^{s_a}$ for check syndrome $s_a$. Normalized min-sum (NMS) scales this magnitude by a factor $0<\alpha\leq1$ to compensate for the overestimation introduced by the MS approximation~\cite{francisco_quantum_min_sum_fpga,maan2026decoding}. Negative output LLRs give hard decisions $\hat e_i=1$, and nonnegative values give $\hat e_i=0$.

GARI~\cite{maan2026decoding} reformulates the correlated detector model $\demxyz$ using transformed variables $\bar e_Z=e_Z+Ue_Y$ and $\bar e_X=e_X+Ve_Y$. The binary matrices $U$ and $V$ satisfy $D_XU=D_X'$ and $D_ZV=D_Z'$, where $D_X'$ and $D_Z'$ describe the detection events produced by $Y$-type errors. Auxiliary checks enforce these transformations, giving the augmented matrix
\begin{equation}
\label{eq:gari-matrix-background}
\demxyzbar=
\begin{pmatrix}
0 & 0 & 0 & \demx & 0\\
0 & 0 & 0 & 0 & \demz\\
I & 0 & U & I & 0\\
0 & I & V & 0 & I
\end{pmatrix},
\end{equation}
with column blocks ordered as $(e_Z,e_X,e_Y,\bar e_Z,\bar e_X)$. The upper blocks impose $D_X\bar e_Z=s_X$ and $D_Z\bar e_X=s_Z$, while the lower blocks have zero syndrome. This construction preserves joint inference of the two Pauli-error components while removing the biclique structures discussed in the introduction.

Here, the GARI decoder denotes NMS decoding on $\demxyzbar$ with a hybrid serial-layered schedule. Each iteration first processes the auxiliary checks in the lower part, using two horizontal layers associated with $U$ and $V$, and then processes the $D_X$ and $D_Z$ checks in the upper part using randomized horizontal serial scheduling.

\section{Methods}
\label{sec:methods}

\garipp combines a single GARI decoder with a second-stage decoder for the residual syndrome, as summarized in Fig.~\ref{fig:decoding_pipeline}. We now describe the first-stage stopping criterion and the construction of the residual decoding problem.

\subsection{Two-stage Pipelined Decoding}
\label{sec:two_stage_decoding}

For each decoding instance, GARI is initialized with prior LLRs derived from the circuit noise model and runs for at most $I_{\max}=7$ iterations, using the schedule described in Sec.~\ref{sec:background}. For the $Z$-memory experiments considered here, let $\hat e_{\mathrm{GARI}}^{(t)}$ denote the hard-decision estimate of the transformed error variables $\bar e_X$ after iteration $t$. These variables include the contributions of $Y$-type errors to the $Z$-detector syndrome. The corresponding residual syndrome is
\begin{equation}
\label{eq:residual_syndrome}
s_{\mathrm{res}}^{(t)}
=s_Z\oplus D_Z\hat e_{\mathrm{GARI}}^{(t)}.
\end{equation}
GARI stops when $s_{\mathrm{res}}^{(t)}=0$, and its estimate is accepted without second-stage decoding. This early stopping criterion requires satisfaction of the $Z$-detector constraints, rather than all checks of the augmented GARI graph.

If the stopping criterion is not reached within the maximum iteration budget, the second stage receives the residual syndrome and LLR-derived weights from an iteration $t_*$, selected according to Sec.~\ref{sec:iteration_selection}. We use Union Find~\cite{Delfosse2021almostlineartime} for surface codes, Hypergraph Union Find ~\cite{wu2025minimum,delfosse2022toward} for BB codes, and Chromobius ~\cite{gidney2023newcircuits,sahay2022mobius} for color codes. In each case, the second stage operates on the $D_Z$ detector model.

For a second-stage error estimate $\hat e_{\mathrm{res}}$, the complete estimate is
\begin{equation}
\label{eq:combined_error}
\hat e_{\mathrm{final}}
=\hat e_{\mathrm{GARI}}^{(t_*)}\oplus\hat e_{\mathrm{res}},
\end{equation}
as shown in Fig.~\ref{fig:decoding_pipeline}. Let $L_D$ map error variables of this detector model to logical-observable flips. The corresponding logical prediction is
\begin{equation}
\label{eq:logical_combination}
\begin{aligned}
\ell_{\mathrm{out}}
&=\ell_{\mathrm{GARI}}^{(t_*)}\oplus\ell_{\mathrm{res}},\\
\ell_{\mathrm{GARI}}^{(t_*)}
&=L_D\hat e_{\mathrm{GARI}}^{(t_*)}.
\end{aligned}
\end{equation}
When an error estimate is returned, $\ell_{\mathrm{res}}=L_D\hat e_{\mathrm{res}}$. 

\subsection{Predictive Pipelining}
\label{sec:iteration_selection}

\begin{figure}[!t]
\centering
\includegraphics[width=0.98\columnwidth]{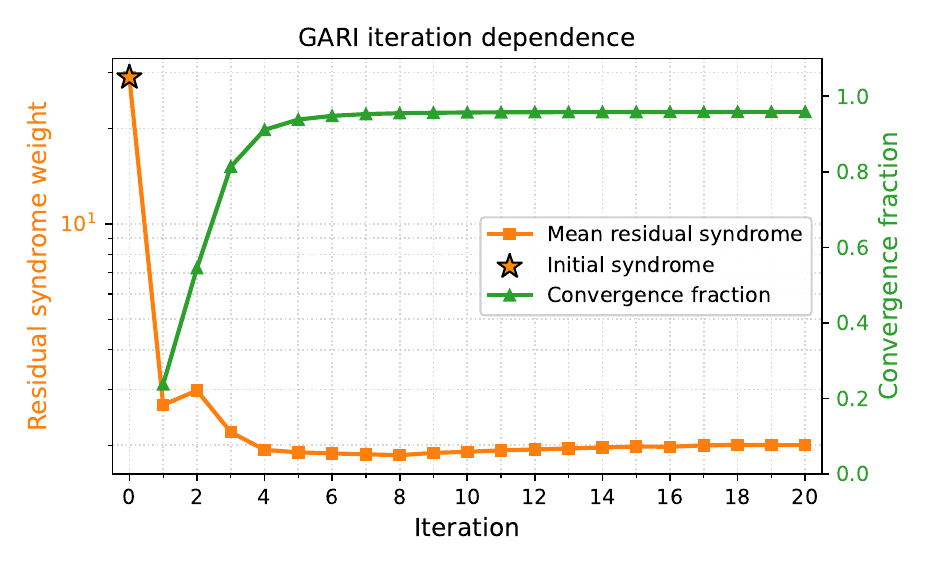}
\caption{Iteration dependence of the residual syndrome weight and convergence fraction for GARI decoding of a distance-$15$ surface code under uniform depolarizing circuit level noise at $p=10^{-3}$, using a maximum of $20$ iterations, and normalization factor $0.875$. Orange squares show the mean residual $Z$-detector syndrome weight among shots still non-converged at each iteration (left logarithmic axis); the star at iteration zero marks the initial mean syndrome weight over nonzero syndrome shots. Green triangles show the convergence fraction (right axis). Iterations are numbered from one after message passing begins. The residual mean reaches a minimum near iteration four and then increases modestly as convergence approaches a plateau. This shows that running the GARI decoder does most of the useful progress in the first few iterations. Thus running the decoder for very large number of iterations and waiting for more progress or using the soft information from the last iteration may not be the best policy as the decoder starts to diverge from the intermediate reliable information. }
    \label{fig:syndrome_weight_vs_itr}
\end{figure}

The extended iteration simulations in Fig.~\ref{fig:syndrome_weight_vs_itr} show that the mean residual syndrome weight among non-converged shots decreases monotonically with iteration number only in the beginning and then either plateaus or begins to increase. This motivates the use of residual syndromes and their corresponding LLRs from intermediate iterations for second stage rather than necessarily using the final iteration.

At iteration $t$, we evaluate the residual syndrome weight
\begin{equation}
\label{eq:residual_weight}
r_t=\lvert s_{\mathrm{res}}^{(t)}\rvert
\end{equation}
and the LLR-magnitude score
\begin{equation}
\label{eq:llr_score}
C^{(t)}
=\sum_{i\in\mathcal I_C}
\left|\mathrm{LLR}_i^{(t)}\right|.
\end{equation}
Here, $\mathcal I_C$ indexes the LLRs associated with the columns of $D_Z$.

The selection criterion uses two quantities: the smallest residual weight encountered in preceding iterations,
\begin{equation}
\label{eq:running_residual_minimum}
m_{t-1}=\min_{u<t}r_u,
\end{equation}
and the score $C_*$ associated with the currently saved estimate. 
The estimate and its associated LLRs are replaced by those from iteration $t$ whenever
\begin{equation}
\label{eq:iteration_update}
r_t<m_{t-1}
\quad\text{or}\quad
C^{(t)}>C_*.
\end{equation}

The running minimum is updated after each iteration. Whenever a new estimate is saved, we set $t_*=t$ and $C_*=C^{(t)}$. Because either condition can trigger replacement, the selected iteration need not minimize the residual syndrome weight but tries to choose a more reliable soft information.

If GARI has not converged after the iteration budget, the second-stage weights are constructed from the selected LLRs as
\begin{equation}
\label{eq:residual_weights}
w_i=\max\!\left(\mathrm{LLR}_i^{(t_*)},0\right).
\end{equation}
The partial GARI estimate is obtained from the LLR signs before clipping. Positive LLRs assign a cost to introducing errors on the variables that GARI estimate as zero, while negative LLRs are assigned zero cost for residual syndrome decoding. Clipping here changes the weights only for the second stage, while the residual syndrome and partial error predication do not change. This can help lower the weight of the final estimated error.

\begin{figure*}[!t]
\centering
\begin{minipage}[t]{0.48\textwidth}
\vspace{0pt}
\centering
\includegraphics[width=0.98\linewidth]{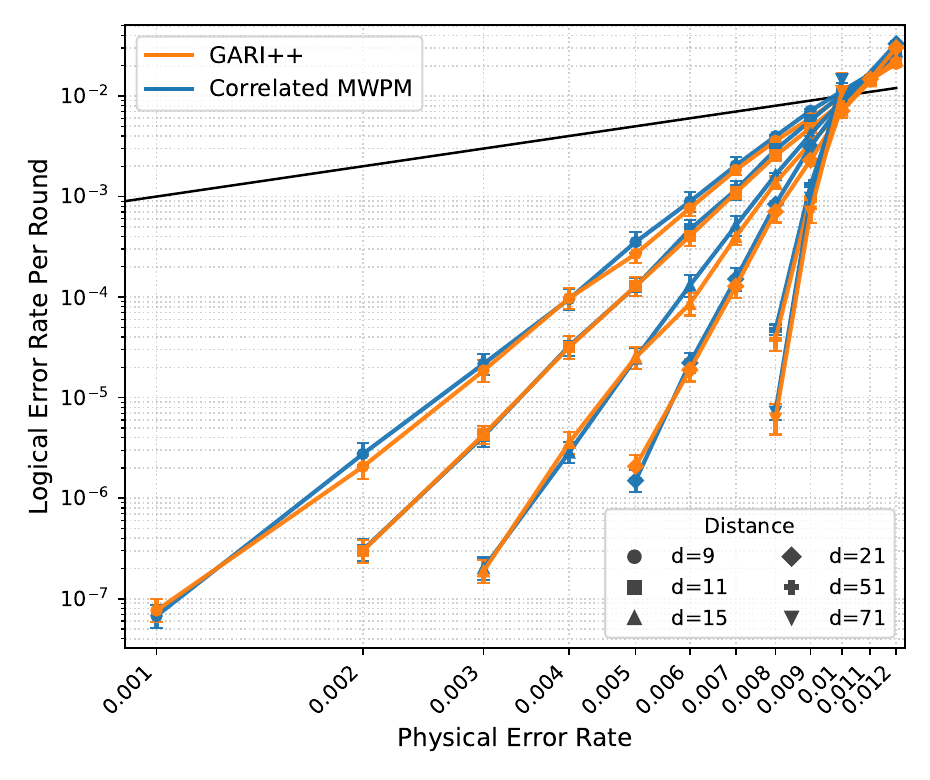}
\caption{Logical error rate per round versus physical error rate for surface code $Z$-memory circuits with uniform depolarising circuit level noise. Blue curves show correlated MWPM and orange curves show \garipp with a residual Union Find. The GARI stage uses a maximum of seven iterations and a normalization factor of 0.9063. The two decoders closely track one another at tested distances upto 71 over the sampled range.
}
\label{fig:accuracy_sc}
\end{minipage}\hfill
\begin{minipage}[t]{0.48\textwidth}
\vspace{0pt}
\centering
\includegraphics[width=0.98\linewidth]{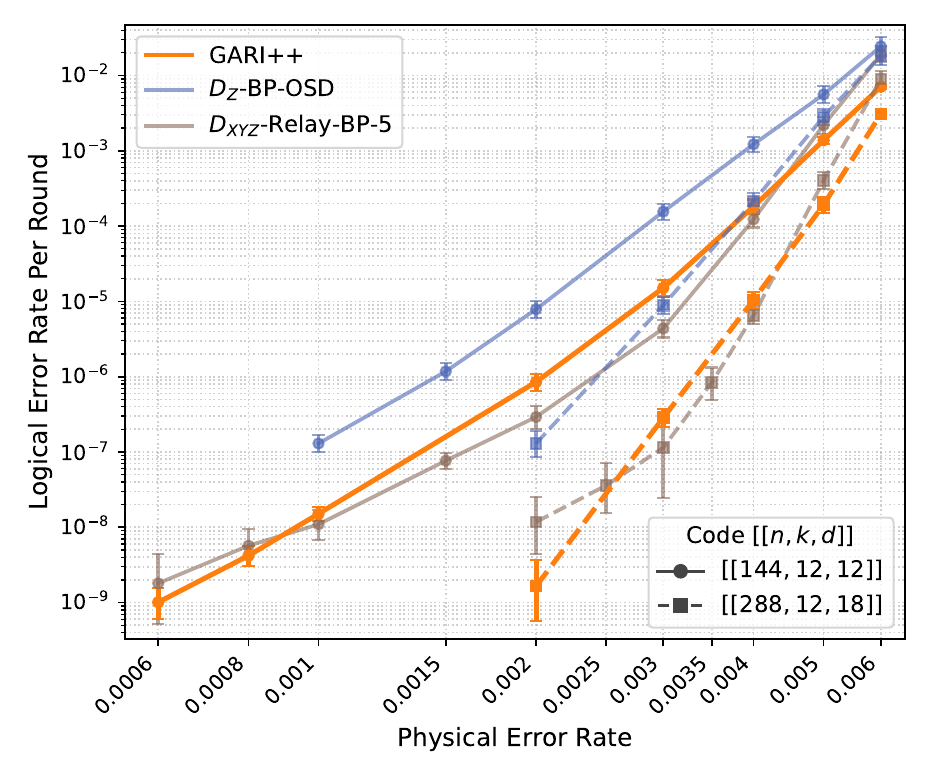}
\caption{Logical error rate per round versus physical error rate for the BB codes $[[144,12,12]]$ and $[[288,12,18]]$ $Z$-memory circuits. \garipp with a residual hypergraph UF (orange) is compared with $\demz$-BP-OSD (blue) and $\demxyz$-Relay-BP-5 (brown). The GARI stage uses a maximum of seven iterations and a normalization factor of 0.725. The extended low noise data show a shallower decrease of the Relay-BP's logical error rates, most clearly for $d=18$, showing the observation of an error floors compared to \garipp}
\label{fig:accuracy_bb}
\end{minipage}
\end{figure*}

\section{Results}
\label{sec:results}

We evaluate how the two outcomes of the \garipp decoding stages combine into logical accuracy, and how syndrome suppression affects the residual syndrome decoding time. Unless explicitly stated otherwise, the reported accuracy, residual syndrome, and timing results use Z memory circuits under uniform depolarizing circuit level noise. Surface code circuits are generated using Stim’s standard rotated surface code construction \cite{Gidney_2021}. For BB codes, we use the circuit implementation of \cite{gong2024lowlatencyiterativedecodingqldpc}. For color codes, we use the superdense construction of \cite{gidney2023newcircuits}, as implemented in the NVIDIA benchmark \cite{olle2026fast,nvidia2026isingdecoding}.
Logical error rates include both GARI converged shots and shots completed by the second stage, and are reported per syndrome extraction round. We compare \garipp with reference decoders for surface, BB codes, and color codes. We also examine residual CPU timing during the residual stage for these codes.

The syndrome weight reduction is accompanied by a substantial reduction in CPU decoding time at larger distances. For surface codes, the mean residual UF decoding time remains below $1\,\mu\mathrm{s}$ per round throughout the investigated distance range on a single Apple M2 CPU core, while the GARI is processed under $1\,\mu\mathrm{s}$ per iteration per round for distances up to 15 on a single xcvu9p FPGA and until distance 25 with $2.6\,\mu\mathrm{s}$ per iteration per round. These results suggest that further architectural optimizations could enable the system to meet even more stringent real-time constraints for code distances up to 25 with off-the-shelf technology.

Each decoding window contains $d$ syndrome extraction rounds. Syndrome weights and CPU decoding times are divided by $d$ to obtain per round quantities. The initial syndrome and residual syndrome averages are evaluated over their respective nonzero input syndromes.

CPU residual decoding times are measured on a single Apple M2 core. The timed operation is the decoding with the detector error model and weights prepared in advance. These measurements exclude GARI decoding, decoder construction, weight preparation and loading, and communication. 

The reported CPU times therefore characterize the second-stage decoding times when invoked, not the end-to-end latency of \garipp.

\subsection{Accuracy}
\label{sec:accuracy}

Figure~\ref{fig:accuracy_sc} shows that \garipp achieves logical accuracy on par with correlated MWPM ~\cite{fowler2013correlated,higgott2025sparse} for surface-code distances up to $71$. The accurate GARI corrections on converged shots are combined with residual syndrome UF decoding of the remaining shots.

% \subsection{Bivariate Bicycle Codes}
% \label{sec:accuracy_bb}

For BB codes, \garipp improves on the $\demz$-BP-OSD reference and achieves lower logical error rates than Relay-BP in the low-noise regime (Fig.~\ref{fig:accuracy_bb}). As the physical error rate decreases, \garipp maintains a consistent suppression of logical errors, whereas the Relay-BP ~\cite{muller2025improvedbeliefpropagationsufficient} curves become shallower. 

\begin{figure}[!t]
\centering
\includegraphics[width=0.98\columnwidth]{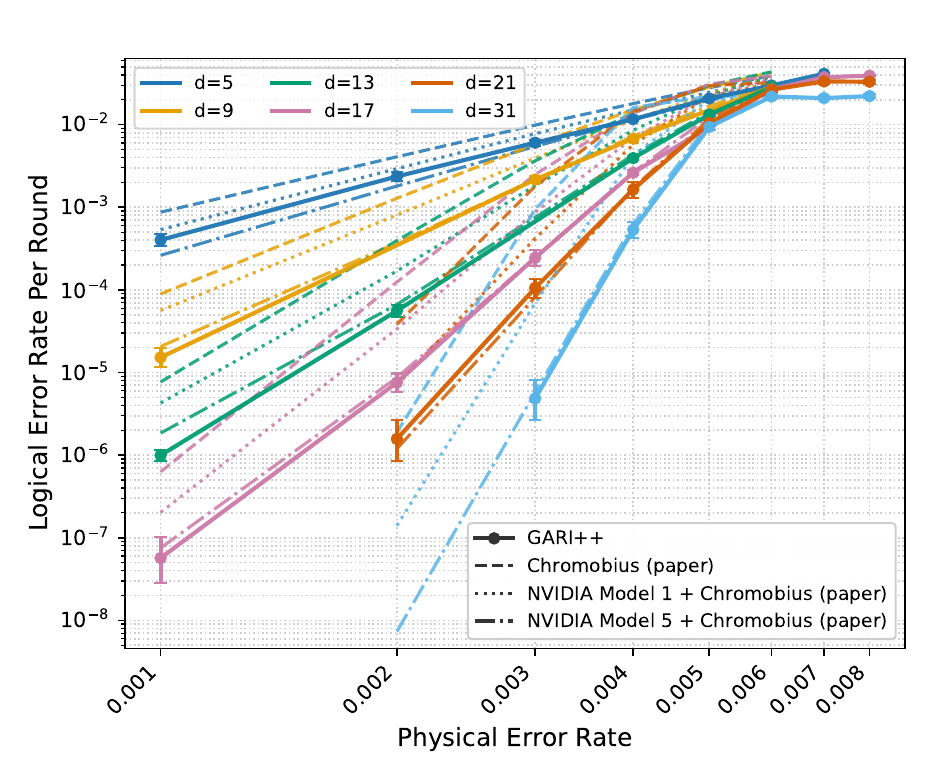}
\caption{Logical error rate per round versus physical error rate for the superdense color code $Z$-memory circuits from ~\cite{olle2026fast} under uniform depolarising circuit level noise. Solid curves with circles show \garipp (with residual Chromobius) with at most seven iterations and normalization factor $0.825$. Dashed, dotted, and dash-dotted curves show the published Chromobius, NVIDIA Model~1 + Chromobius, and NVIDIA Model~5 + Chromobius references, respectively, from Ref.~\cite{olle2026fast}.
}
\label{fig:accuracy_color}
\end{figure}

% \subsection{Color Codes}
% \label{sec:color_code}

Figure~\ref{fig:accuracy_color} shows that \garipp improves on standalone Chromobius, with increasing distance providing stronger logical-error suppression. We apply \garipp to superdense color code ~\cite{gidney2023newcircuits} $Z$-memory circuits from NVIDIA ~\cite{olle2026fast}, using Chromobius as the second stage residual decoder. For the color codes, as well as other codes considered, the residual syndrome per round remains small and decreases across the sampled distances (Fig.~\ref{fig:syndrome_timing}, left axis).

\subsection{Real-Time Decoding}
\label{sec:results_residual}
\label{sec:decoding-latency}

The syndrome suppression in Fig.~\ref{fig:syndrome_timing} is accompanied by a substantial reduction in residual decoding time at larger code distances. The residual syndrome remains small even as the initial syndrome grows, and direct CPU measurements show that the second stage can complete these reduced problems much faster than decoding the original syndrome. 

For surface codes, the mean residual UF decoding time stays below $1\,\mu\mathrm{s}$ per round throughout the investigated range, up to distance $51$. This is below the commonly used superconducting QEC round budget~\cite{muller2025improvedbeliefpropagationsufficient,bascones2026scalable} for the measured decoding mean. Residual HUF likewise reduces the decoding cost for the larger BB codes and residual Chormobius for color codes. These decoding runtimes are distinct from end-to-end latency as the invocation fraction is also needed to determine the average CPU decoding time over all the shots.

Together, these results motivate an FPGA-CPU pipeline: one GARI decoder on programmable logic, followed when needed by residual decoding on an embedded CPU. Prior FPGA implementations~\cite{maan2026decoding,bascones2026scalable} use different GARI configurations. Assessing the complete real-time budget therefore requires matched first-stage timings and the transfer and CPU-processing costs.
\begin{table*}[t]
  \centering
  \caption{Implementation results of GARI for surface codes in an XCVU9P FPGA device. Average iteration counts are for $p=10^{-3}$, with $r=d$ rounds, and non-converged shots contribute the maximum of seven iterations.}
  \label{tab:fpga_results}
  \setlength{\tabcolsep}{3.5pt}
  \renewcommand{\arraystretch}{1.15}
  \resizebox{\textwidth}{!}{%
  \begin{tabular}{|r|r|r|r|r|r|r|r|r|r|r|r|r|r|}
    \hline
    $d$ & Period & $f_{\max}$ & Pipeline & UV & Iter. time & Avg. iter. & LUT & REG & CARRY8 & CLB & Logic & MEM & BRAM \\
        & (ns)   & (MHz)      & stages     & units & $r=d$ ($\mu$s) & $p=0.001$ &     &     &        &     & LUT   & LUT &      \\
    \hline
    25 & 4.08 & 245.28 & 7 & 22 & 63.65 & 4.64 & 189,047 & 343,180 & 1,121 & 47,863 & 188,541 & 506 & 1,397.5 \\
    21 & 3.86 & 259.00 & 7 & 13 & 35.71 & 4.01 &  94,604 & 173,256 &  765 & 26,546 &  94,237 & 367 &  898.5 \\
    19 & 3.91 & 255.62 & 7 & 12 & 26.80 & 3.55 & 77,887 &  130,855 & 673 & 22,128 & 77,530 & 357 & 685\\
    15 & 3.33 & 300.12 & 7 & 13 & 11.23 & 2.65 &  74,671 &  90,531 &  755 & 17,060 &  74,306 & 365 &  472.5 \\
    11 & 3.23 & 309.98 & 7 & 10 &  4.29 & 1.84 &  49,244 &  47,561 &  583 & 10,737 &  48,938 & 306 &  303.5 \\
     9 & 3.22 & 310.66 & 7 &  9 &  2.35 & 1.51 &  44,035 &  37505 &  553 &  8,236 &  43,735 & 300 &  288   \\
     7 & 3.21 & 312.01 & 7 &  9 &  1.11 & 1.28 &  40,012 &  27470 &  531 &  7,807 &  39,724 & 288 &  277   \\
    \hline
  \end{tabular}%
  }
\end{table*}
Building on the architecture presented in~\cite{bascones2026scalable}, surface-code decoders for code distances ranging from 7 to 25 have been implemented on a single xcvu9p FPGA. The pipeline of the architecture and the number of processing units implemented to compute the $U$ or $V$ messages have been adapted to minimize the number of stall clock cycles and optimize latency. This evaluation aims to characterize the scalability of the architecture in terms of area and latency, identify the implementation limits imposed by current FPGA resources, and determine the latency per iteration of the GARI decoder.

The results in Table~\ref{tab:fpga_results} show that the latency per iteration per round for GARI remains below $750 ns$
 for code distances up to 15, while the maximum operating frequency remains nearly constant across these implementations. Combined with the average iterations (see Table~\ref{tab:fpga_results}) used by GARI 0.001 physical error rate , this results in average latency per iteration per round ranging from $201 ns$ to $11.8\mu s$ for surface code distances between 7 and 25.  For distances 21 and 25, the key result is that both implementations fit on a single FPGA without requiring architectural partitioning. However, the increased resource utilization affects timing performance. The maximum operating frequency drops by approximately $18\%$ relative to the distance 15 implementation, resulting in iteration latencies of $1.7\mu s$ and $2.55\mu s$, respectively. These results establish distance 25 as the current implementation limit for the evaluated architecture on the xcvu9p. The observed degradation highlights two distinct resource constraints. First, routing congestion becomes increasingly significant at larger code distances. In particular, the early-stopping criterion unit, which is implemented in parallel, accounts for a substantial share of routing resource consumption in the distance 21 and 25 implementations. Optimizing this unit and its associated interconnections could therefore alleviate routing congestion and improve timing performance. Second, memory capacity becomes an increasingly important constraint. The distance 25 implementation uses more than $50\%$ of the available BRAM resources. This motivates further optimization of message storage and memory organization to reduce BRAM consumption and leave more resources available for scaling to larger code distances.
 Despite these constraints, the results represent a promising first step towards scaling the architecture to larger surface codes. Further architectural optimizations could improve both resource utilization and latency. In addition, larger FPGAs, such as the xcvp1802, which provides substantially greater logic capacity than the xcvu9p, could be explored to assess whether sub-microsecond iteration latency can be achieved at larger code distances. However, architectural optimizations may remain necessary, and increasing the device size alone may not be sufficient to achieve the desired latency.

\section{Conclusion and Discussion}
\label{sec:conclusion_and_disscusion}
We have presented \garipp, a two-stage decoding method that combines a single GARI decoder with residual syndrome decoding. \garipp does not rely on decoding ensembles. The approach exploits two complementary properties of GARI: high logical accuracy when it converges, and strong syndrome suppression even when convergence is not reached within a short iteration budget. In the latter case, the second stage decoder completes the correction using the residual syndrome and GARI-derived weights that were pipelined from the first stage.

Under circuit level uniform depolarizing noise, \garipp achieves competitive logical accuracy across surface, bivariate bicycle and color codes. Its surface code performance is comparable to correlated matching, while the tested BB codes exhibit better accuracy than BP-OSD and Relay-BP. For color codes, combining GARI with residual Chromobius decoding improves on standalone Chromobius by an order of magnitude for distance 17 at 0.001 physical error rate.

\garipp enables a hybrid FPGA-CPU architecture. For the first stage, GARI implementations for surface code distances up to 25 fit on a single FPGA. Syndrome suppression, in turn, reduces the computational cost of the second, CPU stage. At a physical error rate of $10^{-3}$, the mean residual syndrome UF decoding time remains below $1\,\mu\mathrm{s}$ per round for surface code distances even up to 51 on a single Apple M2 core, with almost constant runtime scaling. 

These results can be further improved by investigating a more detailed implementation with either UF implementation on FPGA or on the embedded CPU cores on the same FPGA board, thus lowering the communication time between FPGA and CPU. 

% By combining correlation aware inference with efficient decoding of the residual syndromes, \garipp provides a pathway towards high accuracy, scalable and real-time decoding of quantum LDPC codes.

\begin{acknowledgments}

We used AI tools to help develop parts of the software used for simulations, benchmarking and visualisation of the decoding pipeline, as well as to assist with manuscript preparation. The authors developed the key ideas and take full responsibility for the content.

We thank Gilad Kishony for early feedback on some of the results.
The calculations presented above were performed using computer resources within the Aalto University School of Science “Science-IT” project. We acknowledge the EuroHPC Joint Undertaking for awarding this project access to the EuroHPC supercomputer LUMI, hosted by CSC (Finland) and the LUMI consortium through a EuroHPC Regular Access call. The work of F. Garcia-Herrero and D. Báscones acknowledges support from the project PID2023-147059OB-I00 funded by MCIU/AEI/10.13039/501100011033/FEDER, UE. 

\end{acknowledgments}

\section*{Author contributions}
ASM conceived the original method and two-stage decoding pipeline in discussions with AP, FGH, VS. ASM wrote the code and executed the simulations. FGH and DB developed the FPGA implementation with contributions from ASM. All authors contributed to interpreting the results and writing of the manuscript. AP, FGH and VS verified the results.

\bibliographystyle{apsrev4-2}
\bibliography{__main}

@article{delfosse2022toward,
  title={Toward a union-find decoder for quantum LDPC codes},
  author={Delfosse, Nicolas and Londe, Vivien and Beverland, Michael E},
  journal={IEEE Transactions on Information Theory},
  volume={68},
  number={5},
  pages={3187--3199},
  year={2022},
  publisher={IEEE}
}

@misc{olle2026fast,
  title={Fast and accurate {AI}-based pre-decoders for color codes},
  author={Olle, Jan and Chamberland, Christopher and Li, Muyuan and Baratta, Igor},
  year={2026},
  eprint={2607.10058},
  archivePrefix={arXiv},
  primaryClass={quant-ph},
  url={https://arxiv.org/abs/2607.10058v1}
}

@misc{nvidia2026isingdecoding,
  title={{Ising-Decoding}},
  author={{NVIDIA}},
  year={2026},
  howpublished={GitHub repository},
  note={Color-code circuit-generator snapshot at commit ea7c1bb3b823fc6a38472d0d95974b6d01932288},
  url={https://github.com/NVIDIA/Ising-Decoding/tree/ea7c1bb3b823fc6a38472d0d95974b6d01932288}
}

@misc{gidney2023newcircuits,
  title={New circuits and an open source decoder for the color code},
  author={Gidney, Craig and Jones, Cody},
  year={2023},
  eprint={2312.08813},
  archivePrefix={arXiv},
  primaryClass={quant-ph},
  url={https://arxiv.org/abs/2312.08813}
}

@misc{bascones2026scalable,
  title={{A Scalable FPGA Architecture for Real-Time Decoding of Quantum LDPC Codes Using GARI}},
  author={B{\'a}scones, Daniel and Maan, Arshpreet Singh and Savin, Valentin and Garcia-Herrero, Francisco},
  year={2026},
  eprint={2605.01035},
  archivePrefix={arXiv},
  primaryClass={quant-ph},
  doi={10.48550/arXiv.2605.01035},
  url={https://arxiv.org/abs/2605.01035}
}

@article{maan2026decoding,
  title={Decoding correlated errors in quantum LDPC codes},
  author={Maan, Arshpreet Singh and Garcia Herrero, Francisco Miguel and Paler, Alexandru and Savin, Valentin},
  journal={Nature Communications},
  year={2026},
  publisher={Nature Publishing Group UK London}
}

@article{demarti2024decoding,
  title={Decoding algorithms for surface codes},
  author={deMarti iOlius, Antonio and Fuentes, Patricio and Or{\'u}s, Rom{\'a}n and Crespo, Pedro M and Martinez, Josu Etxezarreta},
  journal={Quantum},
  volume={8},
  pages={1498},
  year={2024},
  publisher={Verein zur F{\"o}rderung des Open Access Publizierens in den Quantenwissenschaften}
}

@article{du2025blindness,
  title={A blindness property of the Min-Sum decoding for the toric code},
  author={Du Crest, Julien and Mhalla, Mehdi and Savin, Valentin},
  journal={IEEE Journal on Selected Areas in Information Theory},
  year={2025},
  publisher={IEEE}
}

@article{roffe2020decoding,
  title={{Decoding across the quantum low-density parity-check code landscape}},
  author={Roffe, Joschka and White, David R and Burton, Simon and Campbell, Earl},
  journal={Physical Review Research},
  volume={2},
  number={4},
  pages={043423},
  year={2020},
  publisher={APS}
}

@article{raveendran2021trapping,
  title={{Trapping sets of quantum LDPC codes}},
  author={Raveendran, Nithin and Vasi{\'c}, Bane},
  journal={Quantum},
  volume={5},
  pages={562},
  year={2021},
  publisher={Verein zur F{\"o}rderung des Open Access Publizierens in den Quantenwissenschaften}
}

@article{Delfosse2021almostlineartime,
  doi = {10.22331/q-2021-12-02-595},
  url = {https://doi.org/10.22331/q-2021-12-02-595},
  title = {{Almost-linear time decoding algorithm for topological codes}},
  author = {Delfosse, Nicolas and Nickerson, Naomi H.},
  journal = {{Quantum}},
  issn = {2521-327X},
  publisher = {{Verein zur F{\"{o}}rderung des Open Access Publizierens in den Quantenwissenschaften}},
  volume = {5},
  pages = {595},
  month = dec,
  year = {2021}
}

@article{koutsioumpas2025automorphismensembledecodingquantum,
  title={Automorphism Ensemble Decoding of Quantum LDPC Codes},
  author={Koutsioumpas, Stergios and Sayginel, Hasan and Webster, Mark and Browne, Dan E},
  journal={arXiv preprint arXiv:2503.01738},
  year={2025}
}

@article{muller2025improvedbeliefpropagationsufficient,
  title={Improved belief propagation is sufficient for real-time decoding of quantum memory},
  author={M{\"u}ller, Tristan and Alexander, Thomas and Beverland, Michael E and B{\"u}hler, Markus and Johnson, Blake R and Maurer, Thilo and Vandeth, Drew},
  journal={arXiv preprint arXiv:2506.01779},
  year={2025}
}

@article{kschischang2001factor,
  title={{Factor graphs and the sum-product algorithm}},
  author={Kschischang, Frank R and Frey, Brendan J and Loeliger, Hans-Andrea},
  journal={IEEE Transactions on Information Theory},
  volume={47},
  number={2},
  pages={498--519},
  year={2001},
  publisher={IEEE}
}

@article{chen2005reduced,
  title={{Reduced-complexity decoding of LDPC codes}},
  author={Chen, Jinghu and Dholakia, Ajay and Eleftheriou, Evangelos and Fossorier, Marc P. C. and Hu, Xiao-Yu},
  journal={IEEE Transactions on Communications},
  volume={53},
  number={8},
  pages={1288--1299},
  year={2005},
  doi={10.1109/TCOMM.2005.852852},
  publisher={IEEE}
}

@article{derks2024designingfaulttolerantcircuitsusing,
  title={Designing fault-tolerant circuits using detector error models},
  author={Derks, Peter-Jan HS and Townsend-Teague, Alex and Burchards, Ansgar G and Eisert, Jens},
  journal={arXiv preprint arXiv:2407.13826},
  year={2024}
}

@article{Gidney_2021,
   title={Stim: a fast stabilizer circuit simulator},
   volume={5},
   ISSN={2521-327X},
   url={http://dx.doi.org/10.22331/q-2021-07-06-497},
   DOI={10.22331/q-2021-07-06-497},
   journal={Quantum},
   publisher={Verein zur Forderung des Open Access Publizierens in den Quantenwissenschaften},
   author={Gidney, Craig},
   year={2021},
   month=jul, pages={497} }

@article{caune2024demonstrating,
  title={Demonstrating real-time and low-latency quantum error correction with superconducting qubits},
  author={Caune, Laura and Skoric, Luka and Blunt, Nick S and Ruban, Archibald and McDaniel, Jimmy and Valery, Joseph A and Patterson, Andrew D and Gramolin, Alexander V and Majaniemi, Joonas and Barnes, Kenton M and others},
  journal={arXiv preprint arXiv:2410.05202},
  year={2024}
}

@article{bravyi2024high,
  title={High-threshold and low-overhead fault-tolerant quantum memory},
  author={Bravyi, Sergey and Cross, Andrew W and Gambetta, Jay M and Maslov, Dmitri and Rall, Patrick and Yoder, Theodore J},
  journal={Nature},
  volume={627},
  number={8005},
  pages={778--782},
  year={2024},
  publisher={Nature Publishing Group UK London},
  url={https://arxiv.org/abs/2308.07915}
}

@article{google2024quantum,
  title={Quantum error correction below the surface code threshold},
  author={{Google Quantum AI and Collaborators}},
  journal={arXiv preprint arXiv:2408.13687},
  year={2024}
}

@article{ryan2021realization,
  title={Realization of real-time fault-tolerant quantum error correction},
  author={Ryan-Anderson, Ciaran and Bohnet, Justin G and Lee, Kenny and Gresh, Daniel and Hankin, Aaron and Gaebler, JP and Francois, David and Chernoguzov, Alexander and Lucchetti, Dominic and Brown, Natalie C and others},
  journal={Physical Review X},
  volume={11},
  number={4},
  pages={041058},
  year={2021},
  publisher={APS}
}

@article{lacroix2024scaling,
  title={Scaling and logic in the color code on a superconducting quantum processor},
  author={Lacroix, Nathan and Bourassa, Alexandre and Heras, Francisco JH and Zhang, Lei M and Bausch, Johannes and Senior, Andrew W and Edlich, Thomas and Shutty, Noah and Sivak, Volodymyr and Bengtsson, Andreas and others},
  journal={arXiv preprint arXiv:2412.14256},
  year={2024}
}

@inproceedings{panteleev2022asymptotically,
  title={Asymptotically good quantum and locally testable classical {LDPC} codes},
  author={Panteleev, Pavel and Kalachev, Gleb},
  booktitle={Proceedings of the 54th Annual ACM SIGACT Symposium on Theory of Computing},
  pages={375--388},
  year={2022}
}

@inproceedings{leverrier2022quantum,
  title={Quantum {Tanner} codes},
  author={Leverrier, Anthony and Z{\'e}mor, Gilles},
  booktitle={2022 IEEE 63rd Annual Symposium on Foundations of Computer Science (FOCS)},
  pages={872--883},
  year={2022},
  organization={IEEE}
}

@article{panteleev2021degenerate,
  title={Degenerate quantum {LDPC} codes with good finite length performance},
  author={Panteleev, Pavel and Kalachev, Gleb},
  journal={Quantum},
  volume={5},
  pages={585},
  year={2021},
  publisher={Verein zur F{\"o}rderung des Open Access Publizierens in den Quantenwissenschaften},
  url={https://arxiv.org/abs/1904.02703}
}

@article{poulin2008iterative,
  title={On the iterative decoding of sparse quantum codes},
  author={Poulin, David and Chung, Yeojin},
  journal={Quantum Information \& Computation},
  volume={8},
  number={10},
  pages={987--1000},
  year={2008},
  publisher={Rinton Press, Incorporated Paramus, NJ}
}

@article{hillmann2024localized,
  title        = {Localized statistics decoding for quantum low-density parity-check codes},
  author       = {Hillmann, Timo and Berent, Lucas and Quintavalle, Armanda O. and Eisert, Jens and Wille, Robert and Roffe, Joschka},
  journal      = {Nature Communications},
  volume       = {16},
  number       = {1},
  pages        = {8214},
  year         = {2025},
  doi          = {10.1038/s41467-025-63214-7},
  publisher    = {Springer Nature},
  url          = {https://doi.org/10.1038/s41467-025-63214-7}
}

@article{gong2024lowlatencyiterativedecodingqldpc,
  title={Toward low-latency iterative decoding of {QLDPC} codes under circuit-level noise},
  author={Gong, Anqi and Cammerer, Sebastian and Renes, Joseph M},
  journal={arXiv preprint arXiv:2403.18901},
  year={2024}
}

@article{higgott2025sparse,
  title={Sparse blossom: correcting a million errors per core second with minimum-weight matching},
  author={Higgott, Oscar and Gidney, Craig},
  journal={Quantum},
  volume={9},
  pages={1600},
  year={2025},
  publisher={Verein zur F{\"o}rderung des Open Access Publizierens in den Quantenwissenschaften},
  url = {https://arxiv.org/pdf/2303.15933}
}

@article{francisco_quantum_min_sum_fpga,
journal ={{IEEE} Access},
author={Javier Valls Coquillat and Francisco Garcia Herrero and Nithin Raveendran and Bane Vasi\'c},
title={Syndrome-based Min-sum vs {OSD-0}  decoders: {FPGA} Implementation and Analysis for Quantum {LDPC} codes},
month = {Oct.},
volume={9},
number={},
pages={138734-138743},
year = {2021}, 
doi={10.1109/ACCESS.2021.3118544}
}

@article{higgott2023improved,
  title={Improved decoding of circuit noise and fragile boundaries of tailored surface codes},
  author={Higgott, Oscar and Bohdanowicz, Thomas C and Kubica, Aleksander and Flammia, Steven T and Campbell, Earl T},
  journal={Physical Review X},
  volume={13},
  number={3},
  pages={031007},
  year={2023},
  publisher={APS}
}

@article{wu2025minimum,
  title={Minimum-Weight Parity Factor Decoder for Quantum Error Correction},
  author={Wu, Yue and Li, Binghong and Chang, Kathleen and Puri, Shruti and Zhong, Lin},
  journal={arXiv preprint arXiv:2508.04969},
  year={2025}
}

@article{bluvstein2024logical,
  title={Logical quantum processor based on reconfigurable atom arrays},
  author={Bluvstein, Dolev and Evered, Simon J and Geim, Alexandra A and Li, Sophie H and Zhou, Hengyun and Manovitz, Tom and Ebadi, Sepehr and Cain, Madelyn and Kalinowski, Marcin and Hangleiter, Dominik and others},
  journal={Nature},
  volume={626},
  number={7997},
  pages={58--65},
  year={2024},
  publisher={Nature Publishing Group UK London}
}

@article{caune2023belief,
  title={Belief propagation as a partial decoder},
  author={Caune, Laura and Reid, Brendan and Camps, Joan and Campbell, Earl},
  journal={arXiv preprint arXiv:2306.17142},
  year={2023}
}

@misc{fowler2013correlated,
  title={Optimal complexity correction of correlated errors in the surface code},
  author={Fowler, Austin G.},
  year={2013},
  eprint={1310.0863},
  archivePrefix={arXiv},
  primaryClass={quant-ph},
  doi={10.48550/arXiv.1310.0863},
  url={https://arxiv.org/abs/1310.0863}
}

@article{smith2023localpredecoder,
  title={Local Predecoder to Reduce the Bandwidth and Latency of Quantum Error Correction},
  author={Smith, Samuel C. and Brown, Benjamin J. and Bartlett, Stephen D.},
  journal={Physical Review Applied},
  volume={19},
  pages={034050},
  year={2023},
  doi={10.1103/PhysRevApplied.19.034050},
  eprint={2208.04660},
  archivePrefix={arXiv},
  primaryClass={quant-ph},
  url={https://doi.org/10.1103/PhysRevApplied.19.034050}
}

@article{liyanage2024fpga,
  title={{FPGA}-Based Distributed Union-Find Decoder for Surface Codes},
  author={Liyanage, Namitha and Wu, Yue and Tagare, Siona and Zhong, Lin},
  journal={IEEE Transactions on Quantum Engineering},
  volume={5},
  pages={1--18},
  year={2024},
  doi={10.1109/TQE.2024.3467271},
  eprint={2406.08491},
  archivePrefix={arXiv},
  primaryClass={quant-ph},
  url={https://doi.org/10.1109/TQE.2024.3467271}
}

@article{ziad2025local,
  title={Local clustering decoder as a fast and adaptive hardware decoder for the surface code},
  author={Ziad, Abbas B. and Zalawadiya, Ankit and Topal, Canberk and Camps, Joan and Geh{\'e}r, Gy{\"o}rgy P. and Stafford, Matthew P. and Turner, Mark L.},
  journal={Nature Communications},
  volume={16},
  pages={11048},
  year={2025},
  doi={10.1038/s41467-025-66773-x},
  eprint={2411.10343},
  archivePrefix={arXiv},
  primaryClass={quant-ph},
  url={https://doi.org/10.1038/s41467-025-66773-x}
}

@article{barber2025realtime,
  title={A real-time, scalable, fast and resource-efficient decoder for a quantum computer},
  author={Barber, Ben and Barnes, Kenton M. and Bialas, Tomasz and Bu{\u g}dayc{\i}, Okan and Campbell, Earl T. and Gillespie, Neil I. and Johar, Kauser and Rajan, Ram and Richardson, Adam W. and Skoric, Luka and Topal, Canberk and Turner, Mark L. and Ziad, Abbas B.},
  journal={Nature Electronics},
  volume={8},
  number={1},
  pages={84--91},
  year={2025},
  doi={10.1038/s41928-024-01319-5},
  eprint={2309.05558},
  archivePrefix={arXiv},
  primaryClass={quant-ph},
  url={https://doi.org/10.1038/s41928-024-01319-5}
}

@article{sahay2022mobius,
  title={A decoder for the triangular color code by matching on a {M{\"o}bius} strip},
  author={Sahay, Kaavya and Brown, Benjamin J.},
  journal={PRX Quantum},
  volume={3},
  number={1},
  pages={010310},
  year={2022},
  doi={10.1103/PRXQuantum.3.010310},
  eprint={2108.11395},
  archivePrefix={arXiv},
  primaryClass={quant-ph},
  url={https://doi.org/10.1103/PRXQuantum.3.010310}
}

@misc{higgott2026modular,
  title={Low-Overhead Quantum Error Correction with Boundary-Connected Planar Modules},
  author={Higgott, Oscar and Sayginel, Hasan and Heras, Francisco J. H. and He, Zhiyang and Jochym-O'Connor, Tomas and Senior, Andrew W. and Zhang, Lei M. and Edlich, Thomas and Spencer, James S. and McEwen, Matt and Gidney, Craig and Bausch, Johannes and Kohli, Pushmeet and Neven, Hartmut},
  year={2026},
  eprint={2610.03682},
  archivePrefix={arXiv},
  primaryClass={quant-ph},
  doi={10.48550/arXiv.2610.03682},
  url={https://arxiv.org/abs/2610.03682}
}

@misc{bhardwaj2026highrate,
  title={High-rate {qLDPC} processors},
  author={Bhardwaj, Aditya and Ma, Muzhou and Meister, Nadine and King, Robbie and Bluvstein, Dolev and Preskill, John and Cain, Madelyn and Xu, Qian and Huang, Hsin-Yuan},
  year={2026},
  eprint={2607.28795},
  archivePrefix={arXiv},
  primaryClass={quant-ph},
  doi={10.48550/arXiv.2607.28795},
  url={https://arxiv.org/abs/2607.28795}
}

@misc{yang2026gala,
  title={Designer Codes from {GALA}: Compact, Self-Dual, and Rate-1/2 {QEC} on Reconfigurable Atom Arrays},
  author={Yang, Willers and Duckering, Casey and Dua, Arpit},
  year={2026},
  eprint={2608.07431},
  archivePrefix={arXiv},
  primaryClass={quant-ph},
  doi={10.48550/arXiv.2608.07431},
  url={https://arxiv.org/abs/2608.07431}
}

@misc{zhao2026ultrahighrate,
  title={Towards Ultra-High-Rate Quantum Error Correction with Reconfigurable Atom Arrays},
  author={Zhao, Chen and Duckering, Casey and Gu, Andi and Maskara, Nishad and Zhou, Hengyun},
  year={2026},
  eprint={2604.16209},
  archivePrefix={arXiv},
  primaryClass={quant-ph},
  doi={10.48550/arXiv.2604.16209},
  url={https://arxiv.org/abs/2604.16209}
}

\end{document}